\documentclass[twocolumn,prb,superscriptaddress,showpacs]{revtex4}
\usepackage{graphicx}
\usepackage{dcolumn}
\usepackage{bm}
\usepackage{textcomp}

\newcommand{\invcm}{\,cm$^{-1}$}
\newcommand{\microns}{\,$\mu m$}
\begin{document}


\title{Non-adiabatic phonons within the doped graphene layers of XC$_6$ compounds}%

\author{M. P. M. Dean}
\email{mpmd2@cam.ac.uk}
 \affiliation{Cavendish Laboratory, University of Cambridge, JJ Thomson Avenue, Cambridge CB3 0HE, United Kingdom}

\author{C. A. Howard}
    \affiliation{London Centre for Nanotechnology and Department of Physics and Astronomy, University College London, London WC1E 6BT, United Kingdom}

\author{S. S. Saxena}%
 \affiliation{Cavendish Laboratory, University of Cambridge, JJ Thomson Avenue, Cambridge CB3 0HE, United Kingdom}

\author{M. Ellerby}
\affiliation{London Centre for Nanotechnology and Department of Physics and Astronomy, University College London, London WC1E 6BT, United Kingdom}

\date{\today}

\begin{abstract}
We report Raman scattering measurements of BaC$_6$, SrC$_6$, YbC$_6$, and CaC$_6$, which permit a systematic study of the phonons and the electron-phonon interaction within the doped graphene layers of these compounds. The out-of-plane carbon phonon softens as the spacing of the graphene layers is reduced in the series BaC$_6$, SrC$_6$, YbC$_6$, and CaC$_6$. This is due to increasing charge in the $\pi^*$ electronic band. Electron-phonon interaction effects between the in-plane carbon modes at $\approx$ 1500\invcm{} and the $\pi^*$ electrons cause a strong non-adiabatic renormalization. As charge is transferred into the $\pi^*$ band these non-adiabatic effects are found to increase concurrent with a reduction in the phonon lifetime.
\end{abstract}

\pacs{71.20.Tx,63.20.kd,74.25.Kc,78.30.-j}

\keywords{graphite intercalates, electron-phonon coupling, Raman spectroscopy}
\maketitle
\section{Introduction\label{sec:intro}}
There is mounting evidence that the nature of electron-phonon interaction in graphitic systems is not fully understood (e.g.\ in Graphene \cite{Pisana2007,Das2008,Lazzeri2006} or carbon nanotubes (CNTs)\cite{Tsang2007}). In graphite electron-phonon scattering is the main contribution to the phonon lifetime and therefore the Raman linewidth, which is typically 11.5\invcm{}.\cite{Lazzeri2006b} When atoms are inserted into graphite, electrons often act to dope the graphene sheets and the Raman linewidth increases dramatically to $\sim$ 100\invcm{}. The question of the electron-phonon interaction is also interesting in light of the discovery of superconductivity at 11.5\,K in CaC$_6$.\cite{Weller2005,Emery2005a} Measurements of a finite Ca isotope effect \cite{Hinks2007} and of an \emph{s}-wave symmetry superconducting gap\cite{Lamura2006,Sutherland2007} suggests that the superconductivity is electron-phonon mediated. However, several experimental and theoretical techniques implicate different groups of phonons and electrons are responsible for the superconductivity.\cite{Calandra2005,Mazin2005,Calandra2006a,Kim2007,Hinks2007,Sugawara2009,Valla2009}

The adiabatic Born-Oppenheimer approximation, in which electrons are assumed to instantaneously respond to the motion of the ions, has been widely applied within the framework of density functional theory (DFT) to predict the phonon frequencies in metals. \cite{Baroni2001} Despite this approximation not being theoretically justified, the discrepancies introduced are usually small $\lesssim$ 1\,\%.\cite{Saitta2008} However, a much larger effect can be observed provided $|\textbf{\em{q.v}}_F| \ll \omega$, \cite{Engelsberg1963,Saitta2008} where $\textbf{\em{q}}$ is the phonon wavevector, $\textbf{\em{v}}_F$ is the Fermi velocity, and $\omega$ is the phonon frequency. Graphene and materials composed of graphene layers typically have small $\textbf{\em{v}}_F$ perpendicular to the graphene sheets. This makes these systems ideal to observe non-adiabatic effects. Weak non-adiabatic effects have already been observed in metals such as osmium,\cite{Ponosov1998} graphene,\cite{Pisana2007} graphite,\cite{Ishioka2008} and CNTs.\cite{Bushmaker2009} Graphite intercalation compounds (GICs) provide an ideal opportunity to examine phonons within the heavily electron-doped graphene sheets in these compounds, where the increased electronic density of states at the Fermi surface should make these effects more significant.

\begin{table}
\caption{\label{tab:Tcvsd}The values of $d$ and the $T_c$ for the GICs under consideration.\cite{Weller2005,Emery2005a,Kim2007,Nakamae2008}}
\begin{ruledtabular}
\begin{tabular}{lcr}
                   & $d$ (\r{A}) & $T_c$ (K)\\
\hline
BaC$_6$                 & 5.25 & $<$0.08  \\
SrC$_6$                 & 4.95 & 1.65 \\
YbC$_6$                 & 4.57 & 6.5 \\
CaC$_6$                 & 4.52 & 11.5 \\
\end{tabular}
\end{ruledtabular}
\end{table}

Recent theoretical work has shown that for most GICs the completely non-adiabatic zone-center phonon frequencies $\omega^{NA}$ are closer to the experimental values than the adiabatic frequencies $\omega^{A}$.\cite{Saitta2008} Using $\omega^{A}$ and $\omega^{NA}$ along with the experimental phonon frequency $\omega_{BWF}$,  Saitta et al.\ calculate the contribution of the electron-phonon scattering to the phonon full width at half maximum $\gamma^{EPC}_{\sigma}$, which can be compared to the experimental width $\Gamma$. This formalism works quite well for KC$_8$ and CaC$_6$, but the width is overestimated in RbC$_8$. Furthermore LiC$_6$ and KC$_{24}$ fall outside the boundaries of $\omega^{A}$ and $\omega^{NA}$. To the best of our knowledge, we present the first systematic experimental study of these effects within a particular family of GICs and the first published phonon spectra of BaC$_6$, SrC$_6$ and YbC$_6$. The solid experimental trends presented here, and the interpretation thereof, can be applied widely within graphitic materials as the phonons we are measuring are the direct relative of the G mode in graphite, graphene, and CNTs. We choose samples with the same in-plane superlattice: BaC$_6$, SrC$_6$, YbC$_6$, and CaC$_6$. Changing the intercalant within these GICs controls the separation of the graphene sheets $d$, which provides a clear physical parameter linked to the changes in $T_c$ in these compounds. Table \ref{tab:Tcvsd} shows that as $d$ decreases the superconducting transition temperature $T_c$ increases dramatically. We show that reducing $d$, by changing the intercalant, leads to a softening of the C$_z$ phonons modes, which is attributed to charge transfer into the $\pi^*$ states. Using this, we go on to demonstrate that the C$_{xy}$ phonon modes are strongly non-adiabatic, and that the size of these effects increases with increasing charge transfer to the $\pi^*$ states. Furthermore, our work provides new evidence that the size of the electron-phonon interaction is stronger than the current theory predicts.\cite{Saitta2008}

\section{Experimental methods}
Samples were synthesized using the vapor transport method \cite{Weller2005,Kim2007} from natural Madagascan graphite flake. X-ray diffraction confirmed the phase purity of the samples, which were typically $\sim$ 200\microns{} in size. For comparison CaC$_6$ and BaC$_6$ were also synthesized from highly oriented pyrolytic graphite (HOPG) using the Li alloy method.\cite{Pruvost2004} The spectra were measured at room temperature using a Renishaw InVia confocal micro-Raman system, equipped with a 514.5\,nm Argon ion laser, which was focused to $\sim$ 3\microns{} onto the ab-plane of the sample. The laser power at the sample was kept below 4\,mW to avoid laser heating effects.
\begin{figure}
\includegraphics[width=0.5\textwidth]{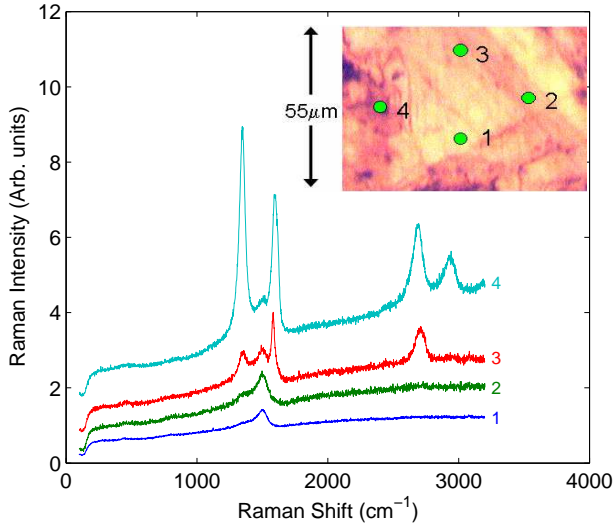}
\caption{\label{fig:surface} (Color online) The Raman spectra from different positions numbered on the GIC surface shown in the increased-contrast inset. Spot 1 corresponds to pure, ordered GIC with one C$_{xy}$ mode visible at $\approx$ 1500\invcm{}. Spots 2-4 also show the $\approx$ 1350\invcm{} and the $\approx$ 1600\invcm{} phonons which arise from sample disorder. All spectra are normalized to represent the same counting time.}
\end{figure}

When dopant atoms are introduced into graphite, the in-plane C$_{xy}$ mode of E$_{2g}$ symmetry (also called the G mode), originally at $\sim$ 1580\invcm{} in graphite and CNT, softens and increases in linewidth. The new superlattice also folds a mode present at the K-point in the graphite Brillouin zone to the $\Gamma$-point, making it Raman active. This C$_z$ mode, around 500\invcm{}, involves out-of-plane C motion.

Alkali and alkali-earth GICs react readily with oxygen and water.\cite{Dresselhaus2002,EnokiBook} The samples were therefore cleaved in an argon glove-box (H$_2$O and O$_2$ levels $<$0.1\,ppm) before being immediately transferred to pre-baked quartz tubes and sealed under a vacuum of $<10^{-6}$\,mbar.

The inset of Fig.\ \ref{fig:surface} shows a typical increased contrast image of the surface of the GICs considered here. While to the naked eye all samples look extremely shiny, under high magnification one can see and avoid step-edges or small areas of surface degradation, which cause the surface to appear slightly blackened. This inhomogeneity explains why Refs.\ \onlinecite{Hlinka2007,Mialitsin2009} observe contaminant peaks in most of their CaC$_6$ spectra. Only by choosing the most reflecting and therefore the most pure part of the sample (position 1 in Fig.\ \ref{fig:surface}) is a clean spectrum obtained containing a C$_{xy}$ mode at $\approx$ 1500\invcm{} and a C$_z$ mode at $\approx$ 450\invcm{} (too weak to see in this plot). We assign, in agreement with Ref.\ \onlinecite{Hlinka2007} the peaks at $\approx$ 1350 and $\approx$ 1600\invcm{} to disordered graphite to which small areas of the surface degrade. These contaminant peaks are inherently much more intense than the true GIC Raman peak, because Raman scattering from disordered graphite is always a resonant process.\cite{Pimenta2007,Ferrari2007}  Ref \onlinecite{Mialitsin2009} attributes the peak at $\approx$ 1350\invcm{} to disordered CaC$_6$. We note that this mode is not associated with the pristine surface.

\section{Results}
\begin{figure}
\includegraphics[width=0.5\textwidth]{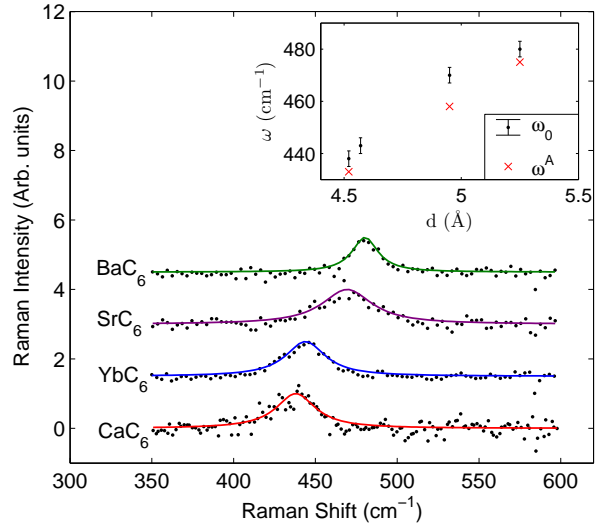}
\caption{\label{fig:puckeringmodes} (Color online) The Raman spectra of the C$_z$ modes for BaC$_6$, SrC$_6$, YbC$_6$, and CaC$_6$. Black dots represent the data points and the solid lines are the Lorentzian fits. A smooth background has been removed from the data. The inset shows the variation in peak position for the experimental frequency $\omega_0$ (dots) and the DFT value $\omega^A$ (crosses) \cite{Calandra2006a}.}
\end{figure}
The C$_z$ modes from BaC$_6$, SrC$_6$, YbC$_6$, and CaC$_6$ are shown in Fig.\ \ref{fig:puckeringmodes}. A clear trend in the peak positions is observed, with the peak softening from 480(3)\invcm{} in BaC$_6$, 470(3)\invcm{} in SrC$_6$, 443(3)\invcm{} in YbC$_6$ to 438(3)\invcm{} in CaC$_6$. To date, experimental phonon measurements of this family of GICs has been limited to CaC$_6$. Out-of-plane phonon dispersions below 50\,meV have been measured by inelastic x-ray scattering \cite{Upton2007}, and found to be close to the DFT predictions. Higher energy modes above 50\,meV have only been measured via Raman scattering, for CaC$_6$ Hlinka et al.\ report $\approx$ 450\invcm{} \cite{Hlinka2007} and Mialitsin et al.\ obtain 440\invcm{}.\cite{Mialitsin2009}
\begin{figure}
\includegraphics[width=0.5\textwidth]{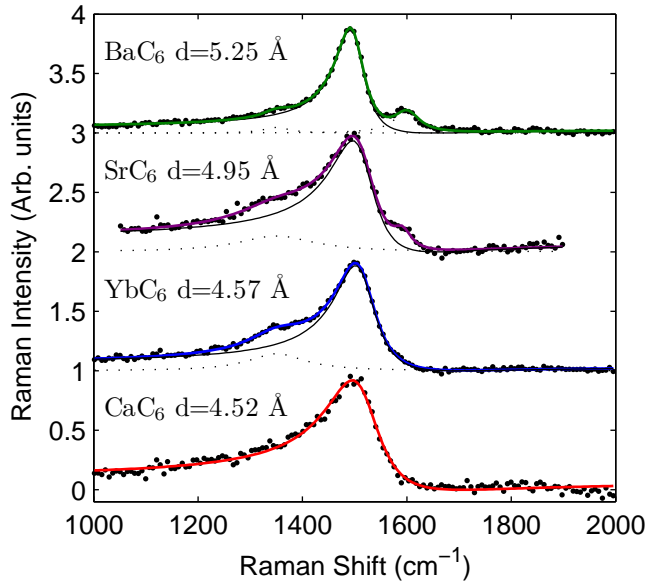}
\caption{\label{fig:BBmodes} (Color online) The Raman spectra of the C$_{xy}$ modes for CaC$_6$, YbC$_6$, SrC$_6$, and BaC$_6$. Black dots are the measured spectrum, the thin line is the BWF function, the dotted lines are the surface contaminant and the fit is shown as a solid line. A linear background has been removed from the data and the spectra have been offset from one another for clarity.}
\end{figure}

The C$_{xy}$ modes in GICs often possess the asymmetric Breit-Wigner-Fano (BWF) line shape, \cite{Nemanich1977,Eklund1979,Dresselhaus2002,EnokiBook} which is modeled as a Raman signal intensity of
\begin{equation}
I(\omega)=I_0 \frac{(1+\frac{\omega-\omega_{BWF}}{q\Gamma/2})^2} {1+(\frac{\omega-\omega_{BWF}}{\Gamma/2})^2} \label{BWFeqn}
\end{equation}
where $1/q$ quantifies the asymmetry of the shape and $\omega_{BWF}$ and $\Gamma$ are fitting parameters to the central frequency and full width at half maximum respectively. In the limit $1/q \rightarrow 0$ a Lorentzian line shape is recovered with a full width $\Gamma$ and frequency $\omega_{0}$. Typical Raman spectra of the C$_{xy}$ modes for the samples considered here are shown in Fig.\ \ref{fig:BBmodes}. The BWF line shape (Eq. \ref{BWFeqn}) was used to fit the peak and Lorentzian functions were also included in some fits to account for any small contamination from the D and G modes. The errors were estimated by fitting a number of different spectra. Table \ref{tab:BBmodes} provides a list of the fitting parameters. The data for CaC$_6$ are in excellent agreement with Hlinka et al.\ who report $\omega_{BWF}$=1511\invcm{}, $\Gamma$=111\invcm{} and $1/q=0.28$.\cite{Hlinka2007} Mialitsin et al.\ measure a spectrum at 20\,K showing large D and G peaks.\cite{Mialitsin2009} They describe a Lorentzian line shape with $\omega_{0}$=1508\invcm{} and $\Gamma$=76\invcm{}.

\section{Analysis and discussion of results}
\begin{table}
\caption{\label{tab:BBmodes}The fit parameters for the C$_{xy}$ mode of the GICs considered here. Values in columns  $\omega^A$ and $\omega^{NA}$ correspond to the DFT calculations of Ref.\ \onlinecite{Saitta2008}. All values except $1/q$ are in \invcm{}.}
\begin{ruledtabular}
\begin{tabular}{lccccccr}
                   & $\omega_{BWF}$ & $\Gamma$ & $1/q$ & $\omega^A$ & $\omega^{NA}$ & $\sigma$ & $\gamma^{EPC}_{\sigma}$\\
\hline
BaC$_6$                 & 1510(5) & 70(10)  & -0.25(5) & 1462 & 1521 & 700(200) & 46(8) \\
SrC$_6$                 & 1510(5) & 100(10) & -0.30(5) & 1459 & 1530 & 910(160) & 64(5) \\
YbC$_6$                 & 1515(5) & 100(10) & -0.30(5) & \dots & \dots & \dots  & \dots  \\
CaC$_6$                 & 1510(3) & 120(10) & -0.35(5) & 1446 & 1529 & 790(80) & 70(4) \\
\end{tabular}
\end{ruledtabular}
\end{table}
The trend in the C$_z$ modes to soften from BaC$_6$, SrC$_6$, YbC$_6$, to CaC$_6$  clearly depends on $d$ rather than the intercalant mass. We propose that by increasing $d$ charge is transferred out of the $\pi^*$ band depopulating the Dirac cones. As the $\pi^*$ band is anti-bonding filling it with electrons destabilizes the bonds causing the softening of the phonons, thus these modes are sensitive to charge transfer. Boeri et al.\ predict C$_z$ modes soften considerably when charge is transferred into the $\pi^*$ states in jellium intercalated graphite.\cite{Boeri2007} The inset of Fig.\ \ref{fig:puckeringmodes} compares the peak values to adiabatic DFT calculations.\cite{Calandra2006a} Theory predicts that the difference between $\omega^A$ and $\omega^{NA}$ is due to intra-band electron-phonon scattering, as inter-band process contribute equally to $\omega^A$ and $\omega^{NA}$.\cite{Lazzeri2006} Given that the experimental values are close to $\omega^A$, and assuming that the $|\textbf{\em{q.v}}_F| \ll \omega$ condition is fulfilled, this agreement shows that there is very little renormalization of these modes due to intra-band transitions. In addition the line-widths of these modes are relatively small and, within the experimental errors, show no trends with $d$. 
The strong doping sensitivity of the C$_z$ modes means that the experimental frequency can be used, in conjunction with DFT, to infer a bulk value of the charge transfer. The experimental value for CaC$_6$ of $438(3) cm^{-1}$ is close to the DFT prediction of 433\invcm{}.\cite{Calandra2006a} In this model 0.20 electron per C atom \footnote{M. Calandra (private communication)} are found in the $\pi^*$ states, which is close to the ARPES derived value of 0.22.\cite{Valla2009} This supports the view that the charge transfer in the surface layers probed in angle-resolved photoemission spectroscopy (ARPES) is representative of the bulk, and shows that these techniques are consistent in this regard. The charge transfer dependence proposed here also explains the series of GICs CsC$_8$, RbC$_8$ to KC$_8$,\cite{Eklund1977} which show a similar trend. The gradient of these effects is similar, but the overall values are different, since in XC$_8$ GICs (where X = intercalant) the charge transfer is less. We also note that in XC$_8$ GICs the C$_z$ mode is folded from $M$ in the graphite Brillouin zone, whereas for IC$_6$ GICs the mode is folded from $K$, which is at slightly lower energy.

The $\omega_{BWF}$ values shown in table \ref{tab:BBmodes} show no dependence on intercalant mass: these phonons only sample the environment of the (doped) graphene planes. These values are all significantly different to the values predicted using DFT \cite{Saitta2008} with the adiabatic approximation $\omega^{A}$ (see Table \ref{tab:BBmodes}). In fact, the values are closer to the completely non-adiabatic limit $\omega^{NA}$, which demonstrates that non-adiabatic effects are essential to understand these phonon energies.  The degree of non-adiabaticity, may be estimated by $\omega_{BWF}-\omega^{A}$ which increases as charge is transferred to the $\pi^*$ states in BaC$_6$, SrC$_6$ to CaC$_6$. Despite the similarity in $\omega_{BWF}$, $\Gamma$ varies significantly increasing in the order BaC$_6$, SrC$_6$ $\approx$ YbC$_6$ to CaC$_6$. This is indicative of a decrease in phonon lifetime as charge is transferred into the $\pi^*$ states. The trends observed when doping the $\pi^*$ states by changing the intercalant atom in GICs are different to those measured in graphene and metallic CNTs at the much lower doping levels accessed by electrostatic gating.\cite{Pisana2007} These systems show an increase in phonon energy and a decrease in $\Gamma$ with electron-doping. However, different effects are present in the high doping case of the GICs because the Kohn anomaly which exists near $\textbf{\em{q}}$ = 0 in graphene \cite{Lazzeri2006} is displaced to finite $\textbf{\em{q}}$ and the $\pi^*$ band is altered and is accompanied by other intercalant-related bands.\cite{Dresselhaus2002,Calandra2005,Calandra2006a,Boeri2007}

Within Saitta et al.'s theory $\omega_{BWF}$ should fall in between the completely adiabatic limit $\omega^A$ and the completely non adiabatic limit $\omega^{NA}$. The electron scattering rate $\sigma=\hbar/\tau$, where $\tau$ is the electron momentum-relaxation time controls were $\omega_{BWF}$ falls in between these limits. For $\sigma$ $\gg$ $\omega^A$ we find $\omega_{BWF}$ $\rightarrow$ $\omega^A$ and for $\sigma$ $\ll$ $\omega^A$ the system becomes fully non-adiabatic and $\omega_{BWF}$ $\rightarrow$ $\omega^{NA}$. We observe no systematic trend in $\sigma$, and we find the values are highly sensitive to any small errors in the experimental and theoretical frequencies. It is therefore difficult to use this method to accurately extract the electron scattering rate. It is however, possible to rearrange Saitta et al.'s equations for derive the electron-phonon scattering induced linewidth
\begin{equation}
\frac{ \gamma_\sigma^{EPC}}{2} \simeq \sqrt{(\omega_{BWF} - \omega^A)(\omega^{NA} - \omega_{BWF})}.
\end{equation}
The derived values of $\gamma^{EPC}_{\sigma}$ account for the trend in $\Gamma$ to increase going from BaC$_6$, SrC$_6$ to CaC$_6$, but significantly underestimate its absolute value. The $\Gamma$ values reported here should be the intrinsic linewidth in these GICs, with a negligible disorder contribution, as these results are reproducible in samples produced from Madagascan flake graphite and HOPG. This is despite HOPG having a crystallite size of $\sim$ 1\microns{}, far smaller than the $\sim$ 100\microns{} in Madagascan flake graphite. An order of magnitude calculation can be performed to estimate the length scales which correspond to the observed $\Gamma$. The C$_{xy}$ modes in these GICs have a typical lifetime of $\omega_{BWF}/\Gamma$ $\approx$ 15 periods and stretch bonds over $\sim$ 5\,\r{A}, which implies the atomic motion is correlated over only $\sim$ 80\,\r{A}. Madagascan flake based CaC$_6$ and YbC$_6$ samples made in the same way as the samples used in this study have ordered regions of $\sim$ 50\microns{}.\footnote{A.\ C.\ Walters (private communication)} Furthermore, the mean distance between scattering centers, as estimated from the low temperature resistivity is $\sim$ 1000\,\r{A}.\cite{Sutherland2007} Disorder broadening should therefore have an insignificant role here. Anharmonic effects are also unlikely to cause such a large linewidth.\cite{Saitta2008} We therefore suggest that theory underestimates the degree of electron-phonon interaction in these compounds.\cite{Saitta2008}

These results also have implications for the debate outlined in the introduction (Sec. \ref{sec:intro}) regarding which phonons and electrons are most relevant for the superconductivity in SrC$_6$, YbC$_6$ and CaC$_6$. The trend uncovered that $T_c$ increases when charge is transferred into the graphene layers provides strong evidence that the C phonons and the $\pi^*$ electrons are involved in the superconductivity. This is very difficult to reconcile with the isotope effect measurements \cite{Hinks2007} and early theory,\cite{Mazin2005} which suggest that superconductivity is predominately due to the intercalant atoms. Our work supports the assertion that the C atoms are relevant, either the C$_z$ modes \cite{Calandra2005,Calandra2006a,Kim2007} or the C$_{xy}$ modes.\cite{Valla2009}

\section{Conclusions}
In conclusion, we present phonon measurements of BaC$_6$, SrC$_6$, YbC$_6$, and CaC$_6$. Raman spectroscopy is employed to monitor the systematic changes in both the phonons and the electron-phonon interaction in the graphene sheets as $d$ is varied by changing the intercalant atom. The C$_z$ phonons are found to soften in the order BaC$_6$, SrC$_6$, YbC$_6$ to CaC$_6$ indicative of change transfer into the $\pi^*$ states as $d$ is reduced. Concurrent with this charge transfer is a large increase in $T_c$. While the out-of-plane carbon mode is well described within the Born-Oppenheimer approximation, the in-plane carbon mode is not. We demonstrate that the in-plane carbon modes show strong non-adiabatic effects, which lead to phonon frequencies that are considerably higher than the adiabatic prediction. As the charge transfer to the $\pi^*$ states increases, these effects get stronger causing an increasing discrepancy between the adiabatic prediction and experimental frequency and an increase in the phonon linewidth. Theoretical work (Ref.\  \onlinecite{Saitta2008}) anticipates the relevance of the non adiabatic effects in these GICs. However, our work provides new evidence that the degree of electron-phonon interaction for this mode is stronger than the current theory predicts. Non-adiabatic effects can be used to extract information on the electrons, phonons and the electron-phonon interaction within the doped graphene layers, which constitute these compounds. Through a systematic study of XC$_6$ GICs, we identify trends which should be widely applicable to characterize a range of graphitic compounds using Raman spectroscopy.
\begin{acknowledgments}
We thank the EPSRC, COST ECOM P16, Selwyn College and Jesus College for funding and the University College London Chemistry Department, G.\ Srinivas, S.\ Firth, and S.\ Albert-Seifried for experimental assistance. We also thank M.\ Calandra, A.\ M.\ Saitta, F.\ Mauri, F.\ Fernandez-Alonso, P.\ B.\ Littlewood, A.\ C.\ Walters, M.\ Sutherland, and  S.\ Bhattacharya for discussions.
\end{acknowledgments}


\begin{thebibliography}{35}
\expandafter\ifx\csname natexlab\endcsname\relax\def\natexlab#1{#1}\fi
\expandafter\ifx\csname bibnamefont\endcsname\relax
  \def\bibnamefont#1{#1}\fi
\expandafter\ifx\csname bibfnamefont\endcsname\relax
  \def\bibfnamefont#1{#1}\fi
\expandafter\ifx\csname citenamefont\endcsname\relax
  \def\citenamefont#1{#1}\fi
\expandafter\ifx\csname url\endcsname\relax
  \def\url#1{\texttt{#1}}\fi
\expandafter\ifx\csname urlprefix\endcsname\relax\def\urlprefix{URL }\fi
\providecommand{\bibinfo}[2]{#2}
\providecommand{\eprint}[2][]{\url{#2}}

\bibitem[{\citenamefont{Pisana et~al.}(2007)\citenamefont{Pisana, Lazzeri,
  Casiraghi, Novoselov, Geim, Ferrari, and Mauri}}]{Pisana2007}
\bibinfo{author}{\bibfnamefont{S.}~\bibnamefont{Pisana}},
  \bibinfo{author}{\bibfnamefont{M.}~\bibnamefont{Lazzeri}},
  \bibinfo{author}{\bibfnamefont{C.}~\bibnamefont{Casiraghi}},
  \bibinfo{author}{\bibfnamefont{K.~S.} \bibnamefont{Novoselov}},
  \bibinfo{author}{\bibfnamefont{A.~K.} \bibnamefont{Geim}},
  \bibinfo{author}{\bibfnamefont{A.~C.} \bibnamefont{Ferrari}},
  \bibnamefont{and} \bibinfo{author}{\bibfnamefont{F.}~\bibnamefont{Mauri}},
  \bibinfo{journal}{Nature Mat.} \textbf{\bibinfo{volume}{6}},
  \bibinfo{pages}{198} (\bibinfo{year}{2007}).

\bibitem[{\citenamefont{Das et~al.}(2008)\citenamefont{Das, Pisana,
  Chakraborty, Piscanec, Saha1, Waghmare, Novoselov, Krishnamurthy, Geim,
  Ferrari et~al.}}]{Das2008}
\bibinfo{author}{\bibfnamefont{A.}~\bibnamefont{Das}},
  \bibinfo{author}{\bibfnamefont{S.}~\bibnamefont{Pisana}},
  \bibinfo{author}{\bibfnamefont{B.}~\bibnamefont{Chakraborty}},
  \bibinfo{author}{\bibfnamefont{S.}~\bibnamefont{Piscanec}},
  \bibinfo{author}{\bibfnamefont{S.~K.} \bibnamefont{Saha1}},
  \bibinfo{author}{\bibfnamefont{U.~V.} \bibnamefont{Waghmare}},
  \bibinfo{author}{\bibfnamefont{K.~S.} \bibnamefont{Novoselov}},
  \bibinfo{author}{\bibfnamefont{H.~R.} \bibnamefont{Krishnamurthy}},
  \bibinfo{author}{\bibfnamefont{A.~K.} \bibnamefont{Geim}},
  \bibinfo{author}{\bibfnamefont{A.~C.} \bibnamefont{Ferrari}},
  \bibnamefont{et~al.}, \bibinfo{journal}{Nature Nano.}
  \textbf{\bibinfo{volume}{3}}, \bibinfo{pages}{210} (\bibinfo{year}{2008}).

\bibitem[{\citenamefont{Lazzeri and Mauri}(2006)}]{Lazzeri2006}
\bibinfo{author}{\bibfnamefont{M.}~\bibnamefont{Lazzeri}} \bibnamefont{and}
  \bibinfo{author}{\bibfnamefont{F.}~\bibnamefont{Mauri}},
  \bibinfo{journal}{Phys. Rev. Lett.} \textbf{\bibinfo{volume}{97}},
  \bibinfo{pages}{266407} (\bibinfo{year}{2006}).

\bibitem[{\citenamefont{Tsang et~al.}(2007)\citenamefont{Tsang, Freitag,
  Perebeinos, Liu, and Avouris}}]{Tsang2007}
\bibinfo{author}{\bibfnamefont{J.~C.} \bibnamefont{Tsang}},
  \bibinfo{author}{\bibfnamefont{M.}~\bibnamefont{Freitag}},
  \bibinfo{author}{\bibfnamefont{V.}~\bibnamefont{Perebeinos}},
  \bibinfo{author}{\bibfnamefont{J.}~\bibnamefont{Liu}}, \bibnamefont{and}
  \bibinfo{author}{\bibfnamefont{P.}~\bibnamefont{Avouris}},
  \bibinfo{journal}{Nature Nanotech.} \textbf{\bibinfo{volume}{2}},
  \bibinfo{pages}{725} (\bibinfo{year}{2007}).

\bibitem[{\citenamefont{Lazzeri et~al.}(2006)\citenamefont{Lazzeri, Piscanec,
  Mauri, Ferrari, and Robertson}}]{Lazzeri2006b}
\bibinfo{author}{\bibfnamefont{M.}~\bibnamefont{Lazzeri}},
  \bibinfo{author}{\bibfnamefont{S.}~\bibnamefont{Piscanec}},
  \bibinfo{author}{\bibfnamefont{F.}~\bibnamefont{Mauri}},
  \bibinfo{author}{\bibfnamefont{A.~C.} \bibnamefont{Ferrari}},
  \bibnamefont{and}
  \bibinfo{author}{\bibfnamefont{J.}~\bibnamefont{Robertson}},
  \bibinfo{journal}{Phys. Rev. B} \textbf{\bibinfo{volume}{73}},
  \bibinfo{pages}{155426} (\bibinfo{year}{2006}).

\bibitem[{\citenamefont{Weller et~al.}(2005)\citenamefont{Weller, Ellerby,
  Saxena, Smith, and Skipper}}]{Weller2005}
\bibinfo{author}{\bibfnamefont{T.~E.} \bibnamefont{Weller}},
  \bibinfo{author}{\bibfnamefont{M.}~\bibnamefont{Ellerby}},
  \bibinfo{author}{\bibfnamefont{S.~S.} \bibnamefont{Saxena}},
  \bibinfo{author}{\bibfnamefont{R.~P.} \bibnamefont{Smith}}, \bibnamefont{and}
  \bibinfo{author}{\bibfnamefont{N.~T.} \bibnamefont{Skipper}},
  \bibinfo{journal}{Nature Phys.} \textbf{\bibinfo{volume}{1}},
  \bibinfo{pages}{39} (\bibinfo{year}{2005}).

\bibitem[{\citenamefont{Emery et~al.}(2005)\citenamefont{Emery, H\'{e}rold,
  d'Astuto, Garcia, Bellin, Mar{\^{e}}ch{\'{e}}, Lagrange, and
  Loupias}}]{Emery2005a}
\bibinfo{author}{\bibfnamefont{N.}~\bibnamefont{Emery}},
  \bibinfo{author}{\bibfnamefont{C.}~\bibnamefont{H\'{e}rold}},
  \bibinfo{author}{\bibfnamefont{M.}~\bibnamefont{d'Astuto}},
  \bibinfo{author}{\bibfnamefont{V.}~\bibnamefont{Garcia}},
  \bibinfo{author}{\bibfnamefont{C.}~\bibnamefont{Bellin}},
  \bibinfo{author}{\bibfnamefont{J.~F.} \bibnamefont{Mar{\^{e}}ch{\'{e}}}},
  \bibinfo{author}{\bibfnamefont{P.}~\bibnamefont{Lagrange}}, \bibnamefont{and}
  \bibinfo{author}{\bibfnamefont{G.}~\bibnamefont{Loupias}},
  \bibinfo{journal}{Phys. Rev. Lett.} \textbf{\bibinfo{volume}{95}},
  \bibinfo{pages}{087003} (\bibinfo{year}{2005}).

\bibitem[{\citenamefont{Hinks et~al.}(2007)\citenamefont{Hinks, Rosenmann,
  Claus, Bailey, and Jorgensen}}]{Hinks2007}
\bibinfo{author}{\bibfnamefont{D.~G.} \bibnamefont{Hinks}},
  \bibinfo{author}{\bibfnamefont{D.}~\bibnamefont{Rosenmann}},
  \bibinfo{author}{\bibfnamefont{H.}~\bibnamefont{Claus}},
  \bibinfo{author}{\bibfnamefont{M.~S.} \bibnamefont{Bailey}},
  \bibnamefont{and} \bibinfo{author}{\bibfnamefont{J.~D.}
  \bibnamefont{Jorgensen}}, \bibinfo{journal}{Phys. Rev. B}
  \textbf{\bibinfo{volume}{75}}, \bibinfo{pages}{014509}
  (\bibinfo{year}{2007}).

\bibitem[{\citenamefont{Lamura et~al.}(2006)\citenamefont{Lamura, Aurino,
  Cifariello, Di Gennaro, Andreone, Emery, H\'{e}rold, Mar\^{e}ch\'{e}, and
  Lagrange}}]{Lamura2006}
\bibinfo{author}{\bibfnamefont{G.}~\bibnamefont{Lamura}},
  \bibinfo{author}{\bibfnamefont{M.}~\bibnamefont{Aurino}},
  \bibinfo{author}{\bibfnamefont{G.}~\bibnamefont{Cifariello}},
  \bibinfo{author}{\bibfnamefont{E.}~\bibnamefont{Di Gennaro}},
  \bibinfo{author}{\bibfnamefont{A.}~\bibnamefont{Andreone}},
  \bibinfo{author}{\bibfnamefont{N.}~\bibnamefont{Emery}},
  \bibinfo{author}{\bibfnamefont{C.}~\bibnamefont{H\'{e}rold}},
  \bibinfo{author}{\bibfnamefont{J.-F.} \bibnamefont{Mar\^{e}ch\'{e}}},
  \bibnamefont{and} \bibinfo{author}{\bibfnamefont{P.}~\bibnamefont{Lagrange}},
  \bibinfo{journal}{Phys. Rev. Lett.} \textbf{\bibinfo{volume}{96}},
  \bibinfo{pages}{107008} (\bibinfo{year}{2006}).

\bibitem[{\citenamefont{Sutherland et~al.}(2007)\citenamefont{Sutherland,
  Doiron-Leyraud, Taillefer, Weller, Ellerby, and Saxena}}]{Sutherland2007}
\bibinfo{author}{\bibfnamefont{M.}~\bibnamefont{Sutherland}},
  \bibinfo{author}{\bibfnamefont{N.}~\bibnamefont{Doiron-Leyraud}},
  \bibinfo{author}{\bibfnamefont{L.}~\bibnamefont{Taillefer}},
  \bibinfo{author}{\bibfnamefont{T.}~\bibnamefont{Weller}},
  \bibinfo{author}{\bibfnamefont{M.}~\bibnamefont{Ellerby}}, \bibnamefont{and}
  \bibinfo{author}{\bibfnamefont{S.~S.} \bibnamefont{Saxena}},
  \bibinfo{journal}{Phys. Rev. Lett.} \textbf{\bibinfo{volume}{98}},
  \bibinfo{pages}{067003} (\bibinfo{year}{2007}).

\bibitem[{\citenamefont{Calandra and Mauri}(2005)}]{Calandra2005}
\bibinfo{author}{\bibfnamefont{M.}~\bibnamefont{Calandra}} \bibnamefont{and}
  \bibinfo{author}{\bibfnamefont{F.}~\bibnamefont{Mauri}},
  \bibinfo{journal}{Phys. Rev. Lett.} \textbf{\bibinfo{volume}{95}},
  \bibinfo{pages}{237002} (\bibinfo{year}{2005}).

\bibitem[{\citenamefont{Mazin}(2005)}]{Mazin2005}
\bibinfo{author}{\bibfnamefont{I.~I.} \bibnamefont{Mazin}},
  \bibinfo{journal}{Phys. Rev. Lett.} \textbf{\bibinfo{volume}{95}},
  \bibinfo{pages}{227001} (\bibinfo{year}{2005}).

\bibitem[{\citenamefont{Calandra and Mauri}(2006)}]{Calandra2006a}
\bibinfo{author}{\bibfnamefont{M.}~\bibnamefont{Calandra}} \bibnamefont{and}
  \bibinfo{author}{\bibfnamefont{F.}~\bibnamefont{Mauri}},
  \bibinfo{journal}{Phys. Rev. B} \textbf{\bibinfo{volume}{74}},
  \bibinfo{pages}{094507} (\bibinfo{year}{2006}).

\bibitem[{\citenamefont{Kim et~al.}(2007)\citenamefont{Kim, Boeri, O'Brien,
  Razavi, and Kremer}}]{Kim2007}
\bibinfo{author}{\bibfnamefont{J.~S.} \bibnamefont{Kim}},
  \bibinfo{author}{\bibfnamefont{L.}~\bibnamefont{Boeri}},
  \bibinfo{author}{\bibfnamefont{J.~R.} \bibnamefont{O'Brien}},
  \bibinfo{author}{\bibfnamefont{F.~S.} \bibnamefont{Razavi}},
  \bibnamefont{and} \bibinfo{author}{\bibfnamefont{R.~K.}
  \bibnamefont{Kremer}}, \bibinfo{journal}{Phys. Rev. Lett.}
  \textbf{\bibinfo{volume}{99}}, \bibinfo{pages}{027001}
  (\bibinfo{year}{2007}).

\bibitem[{\citenamefont{Sugawara et~al.}(2009)\citenamefont{Sugawara, Sato, and
  Takahashi}}]{Sugawara2009}
\bibinfo{author}{\bibfnamefont{K.}~\bibnamefont{Sugawara}},
  \bibinfo{author}{\bibfnamefont{T.}~\bibnamefont{Sato}}, \bibnamefont{and}
  \bibinfo{author}{\bibfnamefont{T.}~\bibnamefont{Takahashi}},
  \bibinfo{journal}{Nature Phys.} \textbf{\bibinfo{volume}{5}},
  \bibinfo{pages}{40} (\bibinfo{year}{2009}).

\bibitem[{\citenamefont{Valla et~al.}(2009)\citenamefont{Valla, Camacho, Pan,
  Fedorov, Walters, Howard, and Ellerby}}]{Valla2009}
\bibinfo{author}{\bibfnamefont{T.}~\bibnamefont{Valla}},
  \bibinfo{author}{\bibfnamefont{J.}~\bibnamefont{Camacho}},
  \bibinfo{author}{\bibfnamefont{Z.-H.} \bibnamefont{Pan}},
  \bibinfo{author}{\bibfnamefont{A.~V.} \bibnamefont{Fedorov}},
  \bibinfo{author}{\bibfnamefont{A.~C.} \bibnamefont{Walters}},
  \bibinfo{author}{\bibfnamefont{C.~A.} \bibnamefont{Howard}},
  \bibnamefont{and} \bibinfo{author}{\bibfnamefont{M.}~\bibnamefont{Ellerby}},
  \bibinfo{journal}{Phys. Rev. Lett.} \textbf{\bibinfo{volume}{102}},
  \bibinfo{pages}{107007} (\bibinfo{year}{2009}).

\bibitem[{\citenamefont{Baroni et~al.}(2001)\citenamefont{Baroni, de~Gironcoli,
  Dal~Corso, and Giannozzi}}]{Baroni2001}
\bibinfo{author}{\bibfnamefont{S.}~\bibnamefont{Baroni}},
  \bibinfo{author}{\bibfnamefont{S.}~\bibnamefont{de~Gironcoli}},
  \bibinfo{author}{\bibfnamefont{A.}~\bibnamefont{Dal~Corso}},
  \bibnamefont{and}
  \bibinfo{author}{\bibfnamefont{P.}~\bibnamefont{Giannozzi}},
  \bibinfo{journal}{Rev. Mod. Phys.} \textbf{\bibinfo{volume}{73}},
  \bibinfo{pages}{515} (\bibinfo{year}{2001}).

\bibitem[{\citenamefont{Saitta et~al.}(2008)\citenamefont{Saitta, Lazzeri,
  Calandra, and Mauri}}]{Saitta2008}
\bibinfo{author}{\bibfnamefont{A.~M.} \bibnamefont{Saitta}},
  \bibinfo{author}{\bibfnamefont{M.}~\bibnamefont{Lazzeri}},
  \bibinfo{author}{\bibfnamefont{M.}~\bibnamefont{Calandra}}, \bibnamefont{and}
  \bibinfo{author}{\bibfnamefont{F.}~\bibnamefont{Mauri}},
  \bibinfo{journal}{Phys. Rev. Lett.} \textbf{\bibinfo{volume}{100}},
  \bibinfo{pages}{226401} (\bibinfo{year}{2008}).

\bibitem[{\citenamefont{Engelsberg and Schrieffer}(1963)}]{Engelsberg1963}
\bibinfo{author}{\bibfnamefont{S.}~\bibnamefont{Engelsberg}} \bibnamefont{and}
  \bibinfo{author}{\bibfnamefont{J.~R.} \bibnamefont{Schrieffer}},
  \bibinfo{journal}{Phys. Rev.} \textbf{\bibinfo{volume}{131}},
  \bibinfo{pages}{993} (\bibinfo{year}{1963}).

\bibitem[{\citenamefont{Ponosov et~al.}(1998)\citenamefont{Ponosov, Bolotin,
  Thomsen, and Cardona}}]{Ponosov1998}
\bibinfo{author}{\bibfnamefont{Y.~S.} \bibnamefont{Ponosov}},
  \bibinfo{author}{\bibfnamefont{G.~A.} \bibnamefont{Bolotin}},
  \bibinfo{author}{\bibfnamefont{C.}~\bibnamefont{Thomsen}}, \bibnamefont{and}
  \bibinfo{author}{\bibfnamefont{M.}~\bibnamefont{Cardona}},
  \bibinfo{journal}{Phys. Stat. Sol. (b)} \textbf{\bibinfo{volume}{208}},
  \bibinfo{pages}{257} (\bibinfo{year}{1998}).

\bibitem[{\citenamefont{Ishioka et~al.}(2008)\citenamefont{Ishioka, Hase,
  Kitajima, Wirtz, Rubio, and Petek}}]{Ishioka2008}
\bibinfo{author}{\bibfnamefont{K.}~\bibnamefont{Ishioka}},
  \bibinfo{author}{\bibfnamefont{M.}~\bibnamefont{Hase}},
  \bibinfo{author}{\bibfnamefont{M.}~\bibnamefont{Kitajima}},
  \bibinfo{author}{\bibfnamefont{L.}~\bibnamefont{Wirtz}},
  \bibinfo{author}{\bibfnamefont{A.}~\bibnamefont{Rubio}}, \bibnamefont{and}
  \bibinfo{author}{\bibfnamefont{H.}~\bibnamefont{Petek}},
  \bibinfo{journal}{Phys. Rev.B} \textbf{\bibinfo{volume}{77}},
  \bibinfo{pages}{121402(R)} (\bibinfo{year}{2008}).

\bibitem[{\citenamefont{Bushmaker et~al.}(2009)\citenamefont{Bushmaker,
  Deshpande, Hsieh, Bockrath, and Cronin}}]{Bushmaker2009}
\bibinfo{author}{\bibfnamefont{A.~W.} \bibnamefont{Bushmaker}},
  \bibinfo{author}{\bibfnamefont{V.~V.} \bibnamefont{Deshpande}},
  \bibinfo{author}{\bibfnamefont{S.}~\bibnamefont{Hsieh}},
  \bibinfo{author}{\bibfnamefont{M.}~\bibnamefont{Bockrath}}, \bibnamefont{and}
  \bibinfo{author}{\bibfnamefont{S.~B.} \bibnamefont{Cronin}},
  \bibinfo{journal}{Nano Lett.} \textbf{\bibinfo{volume}{9}},
  \bibinfo{pages}{607} (\bibinfo{year}{2009}).

\bibitem[{\citenamefont{Nakamae et~al.}(2008)\citenamefont{Nakamae, Gauzzi,
  Ladieu, L'Hote, Emery, H\'{e}rold, Mar{\^{e}}ch{\'{e}}, Lagrange, and
  Loupias}}]{Nakamae2008}
\bibinfo{author}{\bibfnamefont{S.}~\bibnamefont{Nakamae}},
  \bibinfo{author}{\bibfnamefont{A.}~\bibnamefont{Gauzzi}},
  \bibinfo{author}{\bibfnamefont{F.}~\bibnamefont{Ladieu}},
  \bibinfo{author}{\bibfnamefont{D.}~\bibnamefont{L'Hote}},
  \bibinfo{author}{\bibfnamefont{N.}~\bibnamefont{Emery}},
  \bibinfo{author}{\bibfnamefont{C.}~\bibnamefont{H\'{e}rold}},
  \bibinfo{author}{\bibfnamefont{J.}~\bibnamefont{Mar{\^{e}}ch{\'{e}}}},
  \bibinfo{author}{\bibfnamefont{P.}~\bibnamefont{Lagrange}}, \bibnamefont{and}
  \bibinfo{author}{\bibfnamefont{G.}~\bibnamefont{Loupias}},
  \bibinfo{journal}{Solid State Commun.} \textbf{\bibinfo{volume}{145}},
  \bibinfo{pages}{493} (\bibinfo{year}{2008}).

\bibitem[{\citenamefont{Pruvost et~al.}(2004)\citenamefont{Pruvost,
  H\'{e}rold, H\'{e}rold, and Lagrangea}}]{Pruvost2004}
\bibinfo{author}{\bibfnamefont{S.}~\bibnamefont{Pruvost}},
  \bibinfo{author}{\bibfnamefont{C.}~\bibnamefont{H\'{e}rold}},
  \bibinfo{author}{\bibfnamefont{A.}~\bibnamefont{H\'{e}rold}},
  \bibnamefont{and}
  \bibinfo{author}{\bibfnamefont{P.}~\bibnamefont{Lagrangea}},
  \bibinfo{journal}{Carbon} \textbf{\bibinfo{volume}{42}}, \bibinfo{pages}{1825
  } (\bibinfo{year}{2004}).

\bibitem[{\citenamefont{Dresselhaus and Dresselhaus}(2002)}]{Dresselhaus2002}
\bibinfo{author}{\bibfnamefont{M.~S.} \bibnamefont{Dresselhaus}}
  \bibnamefont{and}
  \bibinfo{author}{\bibfnamefont{G.}~\bibnamefont{Dresselhaus}},
  \bibinfo{journal}{Adv. Phys.} \textbf{\bibinfo{volume}{51}},
  \bibinfo{pages}{1} (\bibinfo{year}{2002}).

\bibitem[{\citenamefont{Enoki and Endo}(2003)}]{EnokiBook}
\bibinfo{author}{\bibfnamefont{T.}~\bibnamefont{Enoki}} \bibnamefont{and}
  \bibinfo{author}{\bibfnamefont{M.}~\bibnamefont{Endo}},
  \emph{\bibinfo{title}{Graphite Intercalation Compounds and Applications}}
  (\bibinfo{publisher}{Oxford University Press}, \bibinfo{year}{2003}).

\bibitem[{\citenamefont{Hlinka et~al.}(2007)\citenamefont{Hlinka, Gregora,
  Pokorny, H\'{e}rold, Emery, Mar{\^{e}}ch{\'{e}}, and
  Lagrange}}]{Hlinka2007}
\bibinfo{author}{\bibfnamefont{J.}~\bibnamefont{Hlinka}},
  \bibinfo{author}{\bibfnamefont{I.}~\bibnamefont{Gregora}},
  \bibinfo{author}{\bibfnamefont{J.}~\bibnamefont{Pokorny}},
  \bibinfo{author}{\bibfnamefont{C.}~\bibnamefont{H\'{e}rold}},
  \bibinfo{author}{\bibfnamefont{N.}~\bibnamefont{Emery}},
  \bibinfo{author}{\bibfnamefont{J.~F.} \bibnamefont{Mar{\^{e}}ch{\'{e}}}},
  \bibnamefont{and} \bibinfo{author}{\bibfnamefont{P.}~\bibnamefont{Lagrange}},
  \bibinfo{journal}{Phys. Rev. B} \textbf{\bibinfo{volume}{76}},
  \bibinfo{pages}{144512} (\bibinfo{year}{2007}).

\bibitem[{\citenamefont{Mialitsin et~al.}(2009)\citenamefont{Mialitsin, Kim,
  Kremer, and Blumberg}}]{Mialitsin2009}
\bibinfo{author}{\bibfnamefont{A.}~\bibnamefont{Mialitsin}},
  \bibinfo{author}{\bibfnamefont{J.~S.} \bibnamefont{Kim}},
  \bibinfo{author}{\bibfnamefont{R.~K.} \bibnamefont{Kremer}},
  \bibnamefont{and} \bibinfo{author}{\bibfnamefont{G.}~\bibnamefont{Blumberg}},
  \bibinfo{journal}{Phys. Rev. B} \textbf{\bibinfo{volume}{79}},
  \bibinfo{eid}{064503} (\bibinfo{year}{2009}).

\bibitem[{\citenamefont{Pimenta et~al.}(2007)\citenamefont{Pimenta,
  Dresselhaus, Dresselhaus, Can{\c{c}}ado, Jorio, and Saito}}]{Pimenta2007}
\bibinfo{author}{\bibfnamefont{M.~A.} \bibnamefont{Pimenta}},
  \bibinfo{author}{\bibfnamefont{G.}~\bibnamefont{Dresselhaus}},
  \bibinfo{author}{\bibfnamefont{M.~S.} \bibnamefont{Dresselhaus}},
  \bibinfo{author}{\bibfnamefont{L.~G.} \bibnamefont{Can{\c{c}}ado}},
  \bibinfo{author}{\bibfnamefont{A.}~\bibnamefont{Jorio}}, \bibnamefont{and}
  \bibinfo{author}{\bibfnamefont{R.}~\bibnamefont{Saito}},
  \bibinfo{journal}{Phys. Chem. Chem. Phys.} \textbf{\bibinfo{volume}{9}},
  \bibinfo{pages}{1276} (\bibinfo{year}{2007}).

\bibitem[{\citenamefont{Ferrari}(2007)}]{Ferrari2007}
\bibinfo{author}{\bibfnamefont{A.~C.} \bibnamefont{Ferrari}},
  \bibinfo{journal}{Solid State Commun.} \textbf{\bibinfo{volume}{143}},
  \bibinfo{pages}{47} (\bibinfo{year}{2007}).

\bibitem[{\citenamefont{Upton et~al.}(2007)\citenamefont{Upton, Walters,
  Howard, Rahnejat, Ellerby, Hill, McMorrow, Alatas, Leu, and Ku}}]{Upton2007}
\bibinfo{author}{\bibfnamefont{M.~H.} \bibnamefont{Upton}},
  \bibinfo{author}{\bibfnamefont{A.~C.} \bibnamefont{Walters}},
  \bibinfo{author}{\bibfnamefont{C.~A.} \bibnamefont{Howard}},
  \bibinfo{author}{\bibfnamefont{K.~C.} \bibnamefont{Rahnejat}},
  \bibinfo{author}{\bibfnamefont{M.}~\bibnamefont{Ellerby}},
  \bibinfo{author}{\bibfnamefont{J.~P.} \bibnamefont{Hill}},
  \bibinfo{author}{\bibfnamefont{D.~F.} \bibnamefont{McMorrow}},
  \bibinfo{author}{\bibfnamefont{A.}~\bibnamefont{Alatas}},
  \bibinfo{author}{\bibfnamefont{B.~M.} \bibnamefont{Leu}}, \bibnamefont{and}
  \bibinfo{author}{\bibfnamefont{W.}~\bibnamefont{Ku}}, \bibinfo{journal}{Phys.
  Rev. B} \textbf{\bibinfo{volume}{76}}, \bibinfo{pages}{220501(R)}
  (\bibinfo{year}{2007}).

\bibitem[{\citenamefont{Nemanich et~al.}(1976)\citenamefont{Nemanich, Solin,
  and G{\'{u}}erard}}]{Nemanich1977}
\bibinfo{author}{\bibfnamefont{R.~J.} \bibnamefont{Nemanich}},
  \bibinfo{author}{\bibfnamefont{S.~A.} \bibnamefont{Solin}}, \bibnamefont{and}
  \bibinfo{author}{\bibfnamefont{D.}~\bibnamefont{G{\'{u}}erard}},
  \bibinfo{journal}{Phys. Rev. B} \textbf{\bibinfo{volume}{16}},
  \bibinfo{pages}{2965} (\bibinfo{year}{1976}).

\bibitem[{\citenamefont{Eklund and Subbaswamy}(1979)}]{Eklund1979}
\bibinfo{author}{\bibfnamefont{P.~C.} \bibnamefont{Eklund}} \bibnamefont{and}
  \bibinfo{author}{\bibfnamefont{K.~R.} \bibnamefont{Subbaswamy}},
  \bibinfo{journal}{Phys. Rev. B} \textbf{\bibinfo{volume}{20}},
  \bibinfo{pages}{5157} (\bibinfo{year}{1979}).

\bibitem[{\citenamefont{Boeri et~al.}(2007)\citenamefont{Boeri, Bachelet,
  Giantomassi, and Andersen}}]{Boeri2007}
\bibinfo{author}{\bibfnamefont{L.}~\bibnamefont{Boeri}},
  \bibinfo{author}{\bibfnamefont{G.~B.} \bibnamefont{Bachelet}},
  \bibinfo{author}{\bibfnamefont{M.}~\bibnamefont{Giantomassi}},
  \bibnamefont{and} \bibinfo{author}{\bibfnamefont{O.~K.}
  \bibnamefont{Andersen}}, \bibinfo{journal}{Phys. Rev. B}
  \textbf{\bibinfo{volume}{76}}, \bibinfo{pages}{064510}
  (\bibinfo{year}{2007}).

\bibitem[{\citenamefont{Eklund et~al.}(1977)\citenamefont{Eklund, Dresselhaus,
  Dresselhaus, and Fischer}}]{Eklund1977}
\bibinfo{author}{\bibfnamefont{P.~C.} \bibnamefont{Eklund}},
  \bibinfo{author}{\bibfnamefont{G.}~\bibnamefont{Dresselhaus}},
  \bibinfo{author}{\bibfnamefont{M.~S.} \bibnamefont{Dresselhaus}},
  \bibnamefont{and} \bibinfo{author}{\bibfnamefont{J.~E.}
  \bibnamefont{Fischer}}, \bibinfo{journal}{Phys. Rev. B}
  \textbf{\bibinfo{volume}{16}}, \bibinfo{pages}{3330} (\bibinfo{year}{1977}).

\end{thebibliography}
\end{document}